\def\sbm{{\rm erg \, s^{-1} \, cm^{-2} \, arcmin^{-2}}}

\def\etal{{et al. }}

\def\keV{{\rm keV}}

\def\Omat{\Omega_{\rm M}}
\def\Olam{\Omega_{\Lambda}}

\documentstyle[emulateapj]{article}

\begin{document}
\lefthead{VOIT}
\righthead{CONFUSION OF DIFFUSE X-RAY OBJECTS}
\slugcomment{ApJ Letters, submitted 28 September 2000, accepted 5 December 2000}
\title{Confusion of Diffuse Objects in the X-ray Sky}
\author{G. Mark Voit\footnote{Space Telescope Science Institute,
3700 San Martin Drive, Baltimore, MD 21218, voit@stsci.edu}, 
        August E. Evrard\footnote{Departments of Physics and Astronomy, 
University of Michigan, Ann Arbor, MI 48019},
        Greg L. Bryan\footnote{Department of Physics, 
Massachusetts Institute of Technology, 
Cambridge, MA 02139}$^,$\footnote{Hubble Fellow}
} 

\setcounter{footnote}{0}

\begin{abstract}
Most of the baryons in the present-day universe are thought to
reside in intergalactic space at temperatures of $10^{5-7}$~K.
X-ray emission from these baryons contributes a modest
($\sim$10\%) fraction of the $\sim 1$~keV background  
whose prominence within the large-scale cosmic web depends on 
the amount of non-gravitational energy injected into
intergalactic space by supernovae and AGNs.  Here we show that
the virialized regions of groups and clusters cover over a third
of the sky, creating a source-confusion problem that may hinder
X-ray searches for individual intercluster filaments and 
contaminate observations of distant groups.  
\end{abstract}


\section{Introduction}

Most of the baryons in the universe remain undetected.
We believe they exist because primordial nucleosynthesis
predicts a baryonic matter density amounting to a few percent
of the critical density ($\rho_{\rm cr}$), while the baryons 
associated with stars and gas in galaxies make up less than 
half a percent of $\rho_{\rm cr}$ 
(e.g., Fukugita, Hogan, \& Peebles 1998).
Within clusters of galaxies the intergalactic baryons 
are obvious because gravitational compression causes 
them to glow prominently in X-ray light.  Elsewhere 
they are much more difficult to see.  


Simulations indicate that a large proportion of the
universe's baryons currently reside outside of clusters
in the form of diffuse $10^{5-7}$~K gas associated with 
groups and filaments of galaxies (e.g., Cen \& Ostriker 1999;
Dav\'e \etal 2000).  
The low surface brightness of this warm gas has so 
far made it very challenging to study.
Apart from some tantalizing regions of enhanced X-ray 
surface brightness (Wang, Connolly, \& Brunner 1997; 
Kull \& Bohringer 1999; Scharf \etal 2000) and a few 
O~VI absorption features (Tripp, Savage, \& Jenkins 2000) 
we have no positive detections of it.  

Taken together, the intergalactic baryons lying inside and
outside of clusters ought to contribute a non-negligible fraction
of the $\sim 1$~keV X-ray background.  
If these intergalactic gases were heated by purely gravitational 
processes, they would contribute at least 30\% of the observed 
$\sim 1$~keV X-ray background (Pen 1999; Wu, Fabian, \& Nulsen 
1999). However, the point-source contribution 
at these energies is now estimated to be at least 
80\% (Miyaji, Hasinger, \& Schmidt 2000; Mushotzky \etal 2000),
implying that less than 20\% of this background can be truly 
diffuse.  Non-gravitational energy input by supernovae and
perhaps active galactic nuclei can alleviate this discrepancy
because these energy sources raise the entropy of the intergalactic
gas, making it harder to compress and thereby decreasing 
its contribution to the X-ray background.

Analyses of the temperature-luminosity relation of clusters 
and groups likewise suggest that a significant amount of 
non-gravitational heating has occurred 
(Evrard \& Henry 1991, Kaiser 1991).  The energy 
injected per particle appears to be comparable to the potential 
depth of a typical group, which would severely affect 
the properties of groups (e.g., Ponman, Cannon, \& Navarro 1999).
Models in which early energy injection by supernovae preheats
the intergalactic medium can plausibly account for this
non-gravitational heating.  The motivation for detecting 
intergalactic baryons outside of clusters is therefore
twofold:  Detecting this matter would reveal where the 
baryons implied by primordial nucleosynthesis have gone,
and measuring the entropy level of these baryons would
provide a key constraint on supernova energy injection into
intergalactic space.

This paper shows that detecting X-ray emission from intercluster
structures may prove quite challenging because the virial
radii of groups and clusters out to high redshift cover a
significant fraction of the sky.  
Much of the cosmic X-ray surface brightness, once point sources 
are removed, will come from this confused patchwork of groups 
and clusters, making it difficult to identify true intercluster 
gas and to separate group emission from other virialized sources 
along a common line of sight.

\section{Intercluster Mean Free Path}

Among the major sources of uncertainty that could  
complicate observations of the low surface-brightness structures
in which intergalactic baryons reside are other collapsed structures
along the same line of sight.  The probability that any given 
sight line will encounter a virialized structure with $kT \gtrsim
0.5 \, \keV$ turns out to be of order unity.  Here we 
will estimate that probability using a Press-Schechter 
approach.

For simplicity, let us assume that the distribution of primoridal
perturbations is Gaussian with a dispersion $\propto 
M^{-\alpha}$ on mass scale $M$.  The comoving differential 
number density of virialized objects in the universe can then 
be approximated by $dn/d \nu_c = n_0 (\nu_c / \nu_0)^{-1/\alpha} 
e^{-\nu_c^2 / 2}$ (Press \& Schechter 1974).  
In this expression, $\nu_c$ is the 
critical threshold for virialization of a perturbation
at mass scale $M$ in units of standard deviations, $\nu_0$ 
is the value of $\nu_c$ at some fiducial mass scale $M_0$, 
and $n_0 = (2/\pi)^{1/2}(3H_0^2 \Omat / 8 \pi G M_0)$, where 
$H_0 = 100 \, h \, {\rm km \, s^{-1} \, Mpc^{-1}}$ 
is the current value of the Hubble constant and 
$\Omat$ is the matter density of the universe in units of 
the critical density. 

An appropriate cross-section for a virialized object is
$\sigma_{200} = \pi r_{200}^2$, where $r_{200}$ is the radius
within which the mean overdensity is 200 times the critical 
density.  If we define the mass of a virialized object to be the 
mass within $r_{200}$, then $\sigma_{200}(M) = \pi (GM/100 H^2)^{2/3}$,
where $H$ is the Hubble constant at the redshift of interest.
Thus, a mean free path $\lambda_{200}$ between virialized objects 
more massive than $M$ can be defined by
\begin{eqnarray}
   \lambda_{200}^{-1}(M) & = & \int_{\nu_c(M)}^{\infty} \, \sigma_{200} 
                               \frac {dn} {d \nu_c} \, d \nu_c 
                                  \\
             ~           & = & 2^{- p }
                               \sigma_0 n_0 \nu_0^{1/3\alpha}
                                 \Gamma \left[ 1-p ,
                                \nu_c^2(M) / 2 \right] \; \; ,
                                 \nonumber
\end{eqnarray}
where $\Gamma(a,x)$ is an incomplete gamma function, $p = (1+3\alpha)
/6\alpha$, and $\sigma_0 = \sigma_{200} (M_0)$.

In order to relate this equation to observations, we choose 
$T_0 = 5$~keV to be our fiducial temperature and apply the
low-$z$ mass-temperature relation: $kT_X = (8 \, \keV)
(M/10^{15} \, h^{-1} \, M_\odot)^{2/3}$ (Evrard, Metzler, \& Navarro
1996; Horner, Mushotzky, \& Scharf 1999).  The fiducial
mass and cross-section then become $M_0 = 5 \times 10^{14} \,
h^{-1} \, M_\odot$ and $\sigma_0 = 5.2 \, h^{-2} \, {\rm Mpc}^2$.
To find $\nu_0$, we fit the integral temperature function 
$\int_{\nu_c(T_0)}^{\infty} (dn / d \nu_c) \, d\nu_c$ to the number 
density of $> 5$~keV clusters, yielding $\nu_0 \approx 2.8$ (Donahue
\& Voit 1999).
Thus, for $\alpha = 1/4$, which corresponds to a perturbation slope 
in wavenumber space of $n_k = -1.5$, we arrive at
\begin{equation}
  \frac {c} {H_0 \lambda_{200}(M)} = 0.013 \, 
                    \frac {\Gamma [1-p,\nu_c^2(M) /2]} 
                          {\Gamma [1-p,\nu_0^2 /2]} \; \; . 
\end{equation}
In other words, the probability that a line of sight through the 
low-redshift universe will intercept a $> 5$~keV cluster is 
$\sim 10^{-2}$.

Groups with temperatures of $0.5$-$2$~keV are much more common than
rich clusters and correspondingly cover a much higher percentage
of the sky.  To estimate that covering factor, we extrapolate
$c/H_0 \lambda_{200}$ down to temperatures $\sim 0.5 \, {\rm keV}$
assuming $\alpha  \approx 1/4 - 1/6$ ($n_k \approx -1.5$ to $-2.0$).
For these cooler objects, we find $c/H_0 \lambda_{200} \approx 0.6
- 1.3$.  Because this quantity is of order unity, the projected virial 
radii of objects with potential depths $\gtrsim 0.5$~keV ought to
significantly overlap one another.

This calculation is, of course, quite simplistic in that it fails
to account for factors like clustering of virialized objects,
evolution in their comoving number density, the geometry of
the universe, and the known tendency for the Press-Schechter formula
to overestimate the abundance of group-size halos.  Nevertheless, 
it illustrates two important points: (1) observations of distant 
groups will be corrupted at some level by superpositions with other 
virialized objects, and (2) distinguishing true intercluster baryons 
from those within virialized structures may prove to be quite difficult, 
even under ideal observational conditions.

\section{Covering Factor of Simulated Groups and Clusters}

Large-scale cosmological simulations offer a way to assess the
confusion of projected groups more rigorously.  The largest such
simulations performed to date are the Hubble Volume simulations
of the Virgo Consortium, which model comoving cubes $2-3$~Gpc
on a side, sufficient to reproduce lines of sight stretching
most of the way through the observable universe (Evrard 1999;
MacFarland \etal 1999; Frenk et al. 2000).  Because 
of the magnitude of the computational task, the simulations 
are baryon-free, tracking only the gravitationally-driven 
behavior of dark-matter particles.  Here we will show that total 
solid angle covered by the virial radii of groups and clusters 
resolved in these simulations can approach half of the entire sky.

Hubble Volume simulations have been performed for two 
cosmologies: $\tau$CDM ($\Omat = 1.0$, $\Olam = 0.0$) and 
$\Lambda$CDM ($\Omat = 0.3$, $\Olam = 0.7$).  Several
``light-cone'' catalogs, which list the mass and position 
of each object within the past light-cone of a virtual 
observer located within the simulation volume, have been 
compiled from each simulation, and the catalogued objects 
correspond to spherical regions whose mean mass 
density is 200 times the critical density.
Because the masses of individual particles in these simulations 
are $\approx 2 \times 10^{12} \, h^{-1} \, M_\odot$, a 
typical group contains $< 100$ particles.  The catalogs 
list objects down to a lower limit of 12 particles, too few 
to model the internal properties of a group but enough 
to qualify as a significant density concentration.
An angular size for each simulated cluster or group can be
computed from its catalogued mass and redshift.  The coordinate
distance of such an object is $r(z) = \int_0^z c/H(z) \, dz$, where
$H(z) = H_0 [ \Omat (1+z)^3 + \Olam ]^{1/2}$ in a flat universe,
and its physical diameter is $2 r_{200}$.  These quantities
combine to give an angular diameter of $2 r_{200} (1+z) / r(z)$.

The most convenient type of Hubble Volume catalog for the 
purpose of studying projection effects is the octant catalog
which records cluster properties along a light cone viewed
from a corner of the simulation cube.  For $\Lambda$CDM and
$\tau$CDM the limiting redshifts of the octant catalogs are
$z_{\rm max} = 1.45$ and $1.25$, respectively.  Summing
the projected solid angles of clusters out to the limiting
redshift and mass of each catalog, we find that the total
area covered by virialized objects is 46\% of the sky in
the $\Lambda$CDM case and 52\% of the sky in the $\tau$CDM
case.  Figure~1 shows the projected virial radii of clusters 
and groups more massive than $2.7 \times 10^{13} \, M_\odot$
(12 particles) within a typical square degree of the $\Lambda$CDM 
simulation.  Objects with $0 < z < 0.5$ are shown in blue,
those with $0.5 < z < 1.0$ are in green, and those with
$1 < z < 1.45$ are in red.  Significant overlap
is evident in many cases, and accounting for that overlap 
lowers the net sky coverage to 33\% for $\Lambda$CDM and 
35\% for $\tau$CDM.  These figures should be considered lower
limits for two reasons: Clusters and groups more distant than 
the simulations' redshift limits will provide additional coverage,
and comparisions of the Hubble Volume simulations with smaller-scale
ones indicate that the Hubble Volume catalogs are 10-20\%
incomplete at the low-mass end (Jenkins \etal 2000).

Because highly redshifted groups do not contribute much
flux to the $\sim 1$~keV background, the apparent temperatures 
of projected virialized objects are also worth investigating.  
Using an evolving mass-temperature relation to convert from
mass to temperature (Voit 2000) and dividing each temperature
by $1+z$, we can assign an apparent temperature to each cluster
and group in the Hubble Volume catalogs.  Figure~2 shows the
same field as Figure~1, color-coded by apparent temperature.
Objects with $kT/(1+z) > 1 \, \keV$ are shown in blue,
those with $0.5 \, \keV < kT/(1+z) < 1 \, \keV$ in green, 
$0.5 \, \keV < kT/(1+z) < 1 \, \keV$ in red, and
$kT/(1+z) < 0.25 \, \keV$ in magenta.  The majority of the
sky coverage in this catalog evidently comes from objects with
$kT/(1+z) > 0.25 \, \keV$.

\section{Surface Brightness of Virialized Objects}

Even though virialized objects cover a large percentage of 
the sky, much of that solid angle corresponds to low 
surface-brightness emission.  Current estimates of the
mean $\sim 1$~keV surface brightness attributable to 
intergalactic baryons are in the $1-4 \times 10^{-16}
\, {\rm erg \, cm^{-2} \, s^{-1} \, arcmin^{-2}}$ range 
(Wang \& Ye 1996; Wu \etal 1999; Pen 1999; Dav\'e \etal 2000).  
In this section we crudely estimate the surface brightness 
$I_{200}$ of a virialized object at its virial radius and
show that it is probably within an order of magnitude 
of the mean diffuse surface brightness, implying that
these objects blend into the background in the neighborhood
of their virial radii.

X-ray luminous clusters generally have surface brightness
profiles that are adequately represented by the relation
$I_X(b) = I_0 [ 1 + (b/b_c)^2 ]^{-3 \beta + 1/2}$, where
$b$ is the projected angular radius from the center of the 
cluster, $b_c$ is the angular equivalent of the physical core 
radius $r_c$, and the parameter $\beta$ reflects the behavior 
of $I_X$ at large radii (Cavaliere \& Fusco-Femiano 1978).
Integrating this expression for a typical value of $\beta =
2/3$ gives the cluster's X-ray flux: $F_X = 2 \pi b_c^2 I_0$.
For $b(r_{200}) \gg b_c$, we then obtain
\begin{equation}
  I_{200} \approx \frac {L_X} {8 \pi^2 r_{200}^2}
                  \left( \frac {r_c} {r_{200}} \right)
                  (1+z)^{-4} \; \; ,
\end{equation}
where $L_X$ is the X-ray luminosity of the cluster.

To derive a numerical value for $I_{200}$, we adopt the
core radius-luminosity relation of Jones \etal (1998), 
$r_c = (125 \, h^{-1} \, {\rm kpc}) (L_X / 1.2 \times 10^{44} 
\, h^{-2} \, {\rm erg \, s^{-1}})^{0.2}$, and assume
$r_{200} \propto T^{1/2} H^{-1}$.  We also construct a
ROSAT-band luminosity-temperature relation from the 
relations of Markevitch (1998): $L_X = 2.5 \times 10^{43} \,
h^{-2} \, {\rm erg \, s^{-1}} \, (T/2.3 \, \keV)^\zeta$, where
$\zeta = 2.02$ for $T > 2.3 \, \keV$ and $\zeta = 2.64$ for
$T < 2.3 \, \keV$. For the fiducial temperature of 5~keV, 
we obtain $I_{200} \sim 10^{-15} \, {\rm erg \, cm^{-2} 
\, s^{-1} \, arcmin^{-2}} \, (H/H_0)^3 (1+z)^{-4}$, 
very challenging for ROSAT but well within the capabilities 
of $XMM/Newton$ and $Chandra$.  Note also that the $H^3$ 
factor owing to the increased high-$z$ density scale 
tends to offset the $(1+z)^{-4}$ surface-brightness dimming.

On group scales, $I_{200}$ drops by an order of magnitude 
to below the presumptive mean surface brightness of the 
intergalactic medium.  However, this scaling presumes that 
the luminosity-temperature and core radius-luminosity relations 
we have adopted can be extrapolated to $\lesssim 1$~keV.  
Observations of the $L_X$-$T$ relation for groups show that 
this relation in fact steepens at low temperatures and its scatter 
widens (e.g., Helsdon \& Ponman 2000).  
The scaling of $r_c$ is even less secure, 
given the degeneracy between $\beta$ and $r_c$ in fits to 
flattened surface-brightness profiles.

Combining these surface brightness estimates with the Hubble
Volume catalogs, we can also assess the distribution of 
X-ray surface brightness owing to virialized objects.
Figure~\ref{sbrad_sum} illustrates the fractional area 
$\Omega(>I_X)$ covered by diffuse, virialized gas with a 
ROSAT-band surface-brightness greater than $I_X$.  Splitting 
the sample into objects with masses less than $10^{14} \, h^{-1}
\, M_\odot$ and those greater than that value shows that the
low-mass objects dominate the diffuse surface-brightness 
distribution below $10^{-15} \, \sbm$, raising the possibility
that the properties of these lower-mass systems might profitably 
be studied through statistical analyses of the diffuse X-ray 
background, circumventing the difficulties of selecting and 
modelling X-ray groups in an unbiased way.  In practice, however,
such studies will have to contend with background counts 
generated by charged particles, which can be comparable to
the astrophysical background in {\em Chandra} and {\em XMM/Newton} 
observations. 

These crude estimates show that while the surface-brightness 
profiles of rich clusters should dominate the mean background 
emission from other diffuse structures to at least $r_{200}$, 
the same is not necessarily true for groups.  Emission from near
the virial radii of groups could be seriously diluted by other 
emission from unassociated gas along the same line of sight, 
particularly that from other groups seen in projection.  
Numerical simulations accounting for these projection effects 
will be needed to establish the level of this 
contamination.\footnote{While this paper was being refereed,
Croft \etal (2000) released a preprint describing numerical
simulations of the diffuse X-ray background that support our 
conclusions regarding the difficulty of isolating individual
X-ray filaments among the many groups that tile the X-ray sky.}

\section{Summary}

This paper has shown that the projected virial radii of groups 
and clusters cover much of the sky.  Estimates of this covering
factor based on Press-Schechter analysis place the covering
factor of objects $\gtrsim 0.5 \, \keV$ near unity.  Calculations
of the group-cluster covering factor based on the Hubble Volume
simulations corroborate this estimate.  The virial regions of 
simulated objects with $kT/(1+z) > 0.25$ cover at least a third of the 
sky in both $\tau$CDM and $\Lambda$CDM cosmologies, meaning that
projection effects could potentially contaminate X-ray observations
of distant groups.  These same projection effects will also 
complicate the search for emission from warm baryons associated 
with intercluster filaments.  Untangling the large-scale structures
seen in deep X-ray images may therefore require additional information
from optical redshift surveys that trace the skeleton of dark matter
upon which intergalactic baryons are draped.

\acknowledgements 

The authors would like to thank Caleb Scharf and Megan Donahue 
for useful suggestions.  AEE is supported by AST-9803199 and NASA
NAG5-8458.  GLB is supported by NASA through Hubble Fellowship
grant HF-01104.01-98A from the Space Telescope Science Institute,
which is operated under NASA contract NAS6-26555.  This research 
uses data products made available by the Virgo Consortium.


\newpage

\begin{figure}
\plotone{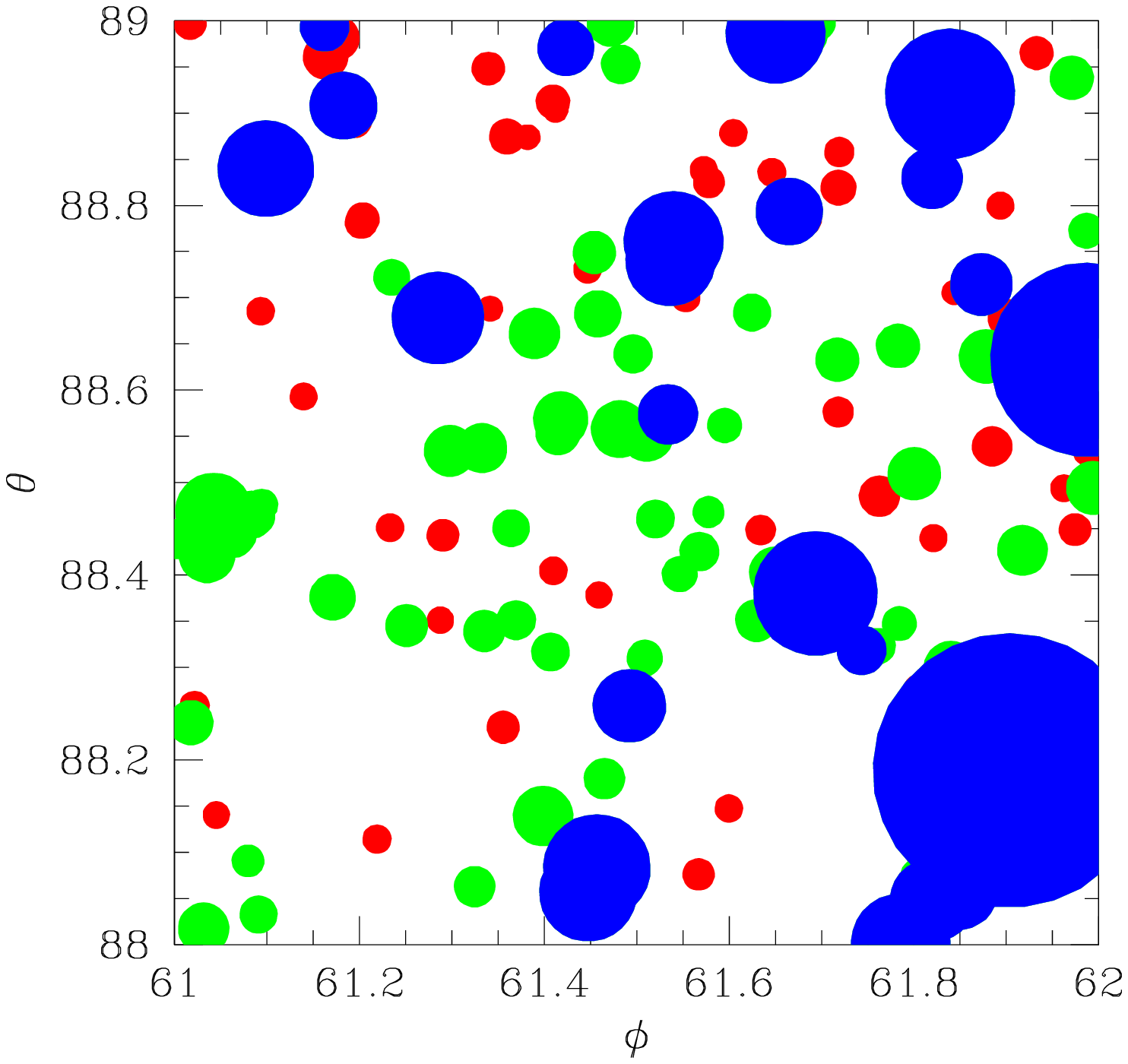}
\figcaption[colcov.ps]{Projected virial radii of simulated clusters
and groups in a $\Lambda$CDM cosmology color-coded by redshift:  
$0 < z < 0.5$ in blue, $0.5 < z < 1$ in green, and $1 < z < 1.45$ 
in red.  The lower mass limit is $2.7 \times 10^{13} \, M_\odot$.
The polar ($\theta$) and azimuthal ($\phi$) coordinates 
of the objects are in units of degrees.  Many virialized objects 
overlap others in the field.  The sum of the solid angles subtended 
by virialized objects is 44\% of the full square degree.
\label{colcov}}
\end{figure}

\begin{figure}
\plotone{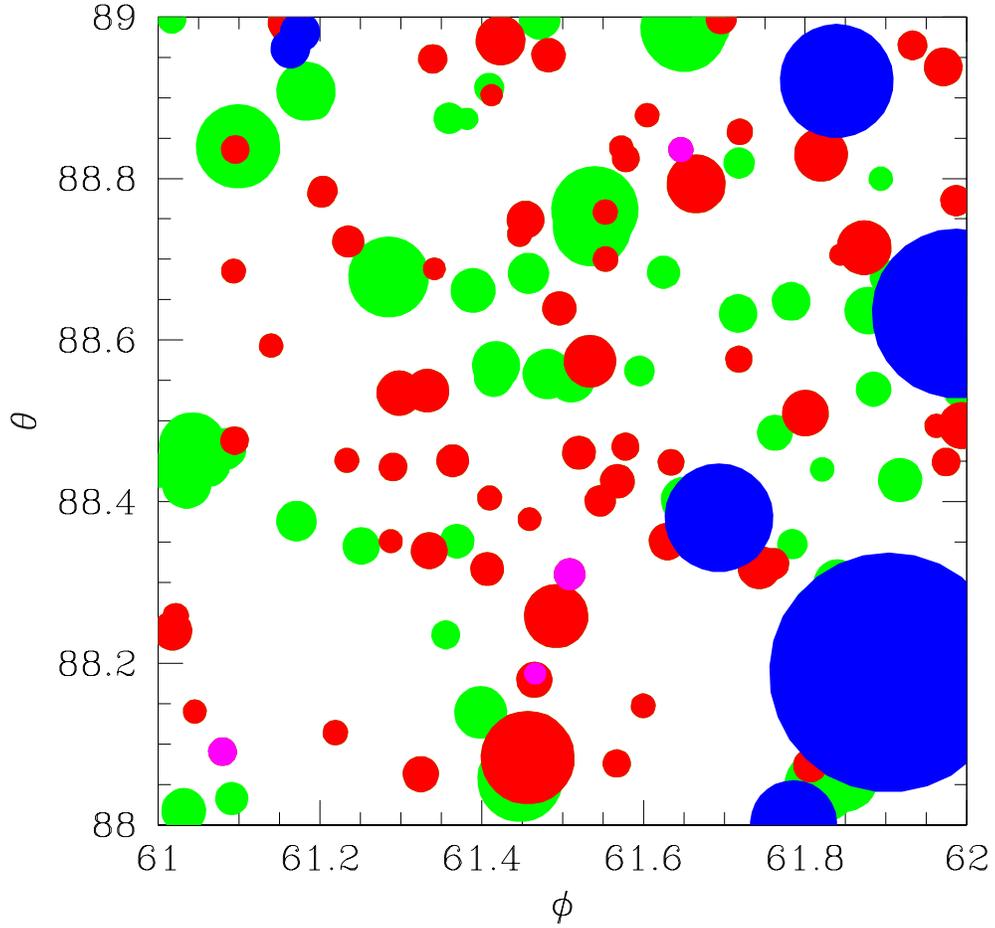}
\figcaption[coltz.ps]{Projected virial radii of simulated clusters
and groups in a $\Lambda$CDM cosmology color-coded by apparent
temperature:  $kT/(1+z) > 1 \, \keV$ in blue, $0.5 < kT/(1+z) < 
1 \, \keV$ in green, $0.25 < kT/(1+z) < 0.5 \, \keV$ in red,
and $0.25 \, \keV < kT/(1+z)$ in magenta.  The lower mass 
limit is $2.7 \times 10^{13} \, M_\odot$, and objects with 
$0.25 < kT/(1+z) < 1 \, \keV$ dominate the sky coverage.
\label{coltz}}
\end{figure}

\begin{figure}
\plotone{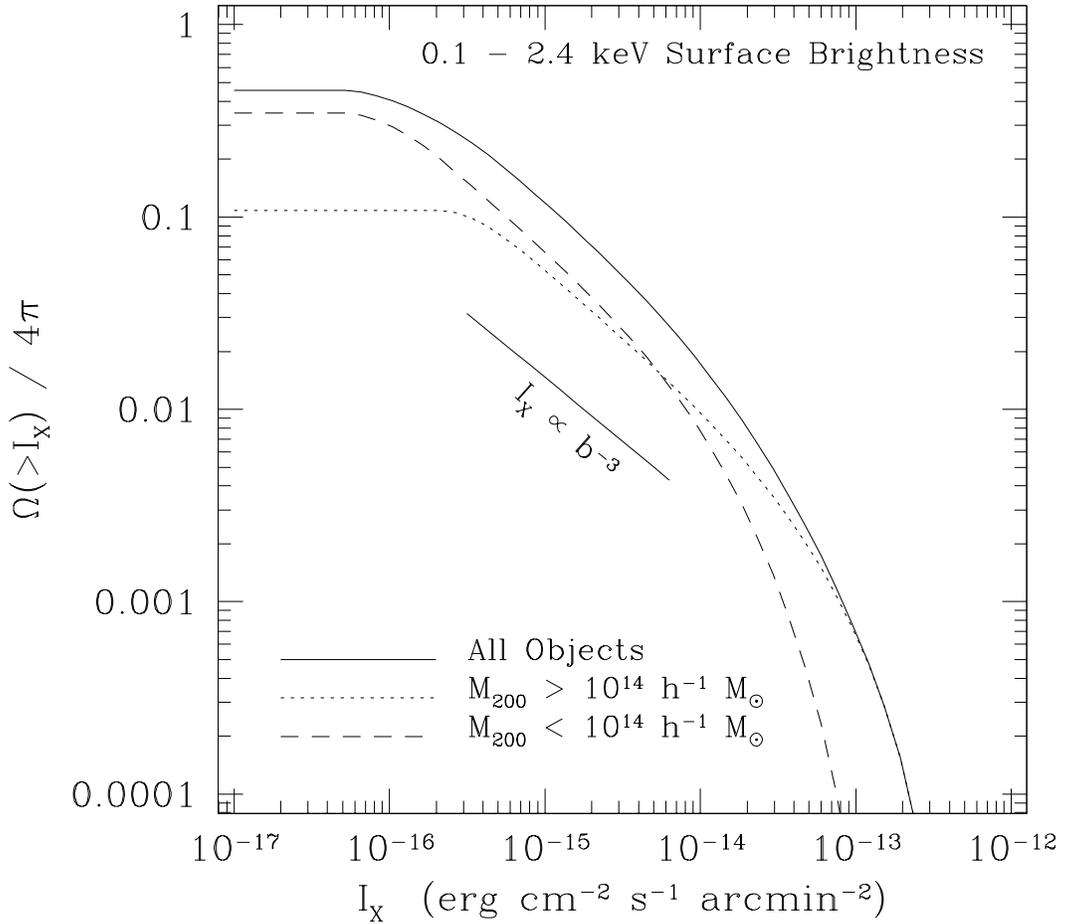}
\figcaption[sbrad_sum.ps]{Total solid angle $\Omega(>I_X)$
covered by virialized regions with 0.1-2.4~keV surface brightness 
greater than $I_X$.  Dotted and dashed lines show the contributions
to $\Omega(>I_X)$ from objects with virial masses ($M_{200}$) 
above and below $10^{14} \, h^{-1} \, M_\odot$, respectively.
The sky coverage levels off at low flux levels because all 
surface-brightness profiles are truncated at $r_{200}$.  
At intermediate values of $I_X$, the slope of $\Omega(>I_X)$
corresponds to a $\beta$-model with $\beta = 2/3$, or equivalently,
$I_X \propto b^{-3}$.
\label{sbrad_sum}}
\end{figure}


\end{document}